\documentclass[12pt]{article}
\usepackage{epsfig}
\usepackage{amsmath,amsfonts,amssymb}

\newcommand{\be}{\begin{equation}}
\newcommand{\ee}{\end{equation}}
\newcommand{\bem}{\begin{displaymath}}
\newcommand{\eem}{\end{displaymath}}
\newcommand{\ba}{\begin{eqnarray}}
\newcommand{\se}{\setcounter{equation}{0}}
\newcommand{\ea}{\end{eqnarray}}
\newcommand{\re}[1]{(\ref{#1})}

\newcommand{\1}{^{-1}}

\newcommand{\delt}{\delta^{\hspace*{-0.2mm}\mbox{\tiny G}}}
\newcommand{\dg}{^{\dagger}}
\newcommand{\di}{\mbox{d}\,}

\newcommand{\f}{{\mbox{\scriptsize f}}} 
\newcommand{\g}{{\mbox{\scriptsize g}}} 
 
\newcommand{\bG}{\bar{G}} 
\newcommand{\G}{{\cal B}} 
\newcommand{\ga}{\gamma_5}
\newcommand{\h}{\frac{1}{2}}
\newcommand{\Id}{\mbox{1\hspace{-1.05mm}l}}

\newcommand{\M}{{\cal M}}

\newcommand{\bP}{\bar{P}} 

\newcommand{\s}{{\cal S}} 
\newcommand{\bs}{\bar{{\cal S}}} 
\newcommand{\ts}{\tilde{{\cal S}}} 
\newcommand{\bS}{\bar{S}} 
\newcommand{\T}{{\cal T}} 
\newcommand{\Tr}{\mbox{Tr}} 
\newcommand{\bu}{\bar{u}} 
\newcommand{\U}{{\cal U}}

\newcommand{\bw}{\bar{w}}

\begin{document}
 
\hfill {\sc HU-EP}-05/30
 
\vspace*{1cm}
 
\begin{center}
 
{\Large \bf Gauge transformations in non-perturbative\\chiral gauge theories}
 
\vspace*{0.9cm}
 
{\bf Werner Kerler}
 
\vspace*{0.3cm}
 
{\sl Institut f\"ur Physik, Humboldt-Universit\"at, D-12489 Berlin,
Germany}
 
\end{center}

\vspace*{1cm} 

\begin{abstract}
We reconsider gauge-transformation properties in chiral gauge theories on 
the lattice observing all pertinent information and show that these
properties are actually determined in a general way for any gauge group 
and for any value of the index. In our investigations we also clarify 
several related issues. 
\end{abstract}

\vspace*{0.3cm}

\section{Introduction}

Gauge invariance of the chiral determinant on the lattice has been considered 
a major problem \cite{go01}. Particular developments aiming at gauge 
invariance have been presented in Ref.~\cite{lu98} and have been the basis
of various further works. Motivated by our observation that actually more 
information is available on the transformation properties of the bases 
involved, we here reinvestigate the subject. We also extend the considerations
beyond the vacuum sector admitting zero modes and any value of the index.

We first show how the expressions for the chiral determinant in 
Ref.~\cite{lu98}, which are based on gauge variations, can be evaluated 
further and find that with the covariance requirement for the current 
introduced there everything is fixed without any further assumptions. It 
thus becomes obvious that the developments in Ref.~\cite{lu98} are not 
relevant for the question of gauge invariance, which we discuss in detail. 

In our more general analysis we then reveal that not allowing arbitrary 
switching to different equivalence classes of pairs of bases is the general
principle which also implies covariance of the current mentioned above. 
To admit zero modes and any value of the index we consider appropriate forms
of fermionic correlation functions and investigate their behavior under 
finite gauge transformations, finding gauge covariance up to constant phase 
factors.

In Section 2 we collect general relations. In Section 3 we consider 
variations of the effective action and the special case of Ref.~\cite{lu98}. 
In Section 4 we use finite transformations to analyze general correlation
functions. Section 5 contains our conclusions.

\section{General relations}\se

\subsection{Basic quantities}

The chiral projections $\bP_+$ and $P_-$ are subject to  
\be
\bP_+D=DP_-,
\label{DP}
\ee
where $D$ is the Dirac operator. They can be expressed as 
\be
P_-=\h(\Id-\ga G),\quad\qquad\bP_+=\h(\Id+\bG\ga),
\label{GaG}
\ee
which because of $P_-\dg=P_-=P_-^2$ and $\bP_+\dg=\bP_+=\bP_+^2$ implies 
$G\1=G\dg=\ga G\ga$ and $\;\bG\1=\bG\dg=\ga\bG\ga$. Requiring $D$ to be 
$\ga$-Hermitian and normal and $G$ and $\bG$ to be functions of $D$ we 
get $\bar{N}-N=I$ for the numbers of anti-Weyl and Weyl degrees of freedom 
$\bar{N}=\Tr\,\bP_+$ and $N=\Tr\,P_-$ and the index $I$ of $D$. 
For more details on the operator properties we refer to 
the recent analysis in Ref.~\cite{ke05}.

Integrating out the Grassmann variables basic fermionic correlation 
functions for the Weyl degrees of freedom are given by \cite{ke03}
\be
\langle\psi_{\sigma_{r+1}}\ldots\psi_{\sigma_N}\bar{\psi}_{\bar{\sigma}_{r+1}}
\ldots\bar{\psi}_{\bar{\sigma}_{\bar{N}}}\rangle_{\f}
=\frac{1}{r!}\sum_{\bar{\sigma}_1\ldots\bar{\sigma}_r}\sum_{\sigma_1,\ldots,
\sigma_r}\bar{\Upsilon}_{\bar{\sigma}_1\ldots\bar{\sigma}_{\bar{N}}}^*
\Upsilon_{\sigma_1\ldots\sigma_N}D_{\bar{\sigma}_1\sigma_1}\ldots
D_{\bar{\sigma}_r\sigma_r}
\label{COR}
\ee
with the alternating multilinear forms
\be
\Upsilon_{\sigma_1\ldots\sigma_N}=\sum_{i_1,\ldots,i_N=1}^N\epsilon_{i_1, 
\ldots,i_N}u_{\sigma_{1}i_{1}}\ldots u_{\sigma_Ni_N},\;\; \bar{\Upsilon}_{
\bar{\sigma}_1\ldots{\bar{\sigma}_{\bar{N}}}}=\sum_{j_1,\ldots,
j_{\bar{N}}=1}^{\bar{N}}\epsilon_{j_1,\ldots,j_{\bar{N}}}\bar{u}_{\bar{
\sigma}_{1}j_{1}}\ldots\bar{u}_{\bar{\sigma}_{\bar{N}}j_{\bar{N}}}.
\label{FO}
\ee
The bases $\bu_{\bar{\sigma}j}$ and $u_{\sigma i}$ in \re{FO} satisfy
\be
P_-=uu\dg,\quad u\dg u=\Id_{\rm w},\qquad\qquad\bP_+=\bu\bu\dg,\quad\bu\dg\bu
=\Id_{\rm\bw}.
\label{uu}
\ee 
General fermionic functions are linear combinations of the basic ones 
\re{COR}.

\subsection{Subsets of bases}

By \re{uu} the bases are only fixed up to unitary transformations, 
$u^{[S]}=uS$, $\bu^{[\bar{S}]}=\bu\bar{S}$, under which the forms \re{FO} 
get multiplied by factors $\det_{\rm w}S$ and $\det_{\rm \bw}\bar{S}$, 
respectively, and therefore the correlation functions \re{COR} by a 
factor\footnote{To compare with
vector theory we can consider its formulation analogous to \re{COR} with 
$\Tr\,\Id$ instead of $\bar{N}$ and $N$ and see that because of $\bu=u$ 
we there get $\,\det S\cdot\det S\dg=1$ instead of \re{DtD}.} 
\be
{\det}_{\rm w}S\cdot{\det}_{\rm\bw}\bar{S}\dg=e^{i\vartheta}.
\label{DtD}
\ee
Therefore, in order that full correlation functions remain invariant, 
we have to require
\be
\vartheta=\mbox{ const},
\label{UNI}
\ee
i.e.~that the phase $\vartheta$ is independent of the gauge field.
While without condition \re{UNI} all bases related to a chiral projection 
are connected by unitary transformations, with it the total set of pairs 
of bases $u$ and $\bu$ is decomposed into inequivalent subsets, beyond 
which legitimate transformations do not connect. This has the important 
consequence that for the formulation of the theory one has to restrict to 
one of such subsets. 

Different ones of the indicated subsets, which obviously are equivalence 
classes, are related by pairs of of basis transformations for which
$\vartheta$ in \re{DtD} depends on the gauge field. The phase factor 
$e^{i\vartheta(U)}$ then determines how the results of the theory differ 
for the respective classes. In view of such differences one has to decide 
which class is appropriate for the description of physics, for which there 
is, however, so far no criterion.

\subsection{Gauge transformations}

Since $G$ and $\bG$ are functions of $D$ the gauge-transformation behavior 
$D'=\T D\T\dg$ is inherited by them and then also by the chiral projections, 
which thus satisfy $P_-'=\T P_-\T\dg$ and $\bP_+'=\T\bP_+\T\dg$ in accordance 
with \re{DP}. In addition to the case where none of the chiral projections
commutes with $\T$ the case where one of them is constant and thus
commutes is of interest (examples of which are the particular choices in 
Ref.~\cite{ke05} and in Ref.~\cite{lu98}, respectively). 

Considering $[\T,P_-]\ne0$ we note that given a solution $u$ of the 
conditions \re{uu}, then $\T u$ is a solution of the transformed conditions 
\re{uu}. To account for the fact that $u$ and $u'$ are only fixed up to
unitary transformations we introduce the unitary transformation $\s$ 
getting $u'=\T u\s$ for all solutions of the transformed conditions. 
Analogous considerations apply to $[\T,\bP_+]\ne0$. In the case where 
$[\T,P_-]\ne0$ and $[\T,\bP_+]\ne0$ we thus have the general relations 
\be
u'=\T u\s,\qquad \bu'=\T\bu\bs.
\label{uTs}
\ee
For the phase $\Theta$ in
\be
{\det}_{\rm w}\s\cdot{\det}_{\rm\bw}\bs\dg=e^{i\Theta(\T)}
\label{D*D}
\ee 
using \re{DtD} with \re{UNI} we then immediately get
\be
\Theta(\Id)=\mbox{const},
\label{THE}
\ee
i.e.~independence of the gauge field at least for $\T=\Id$.

In the case where $[\T,P_-]\ne0$ and $\bP_+=$ const the 
equivalence class of pairs of bases always contains constant $\bu_{\rm c}$. 
This follows since given a pair $u$, $\bu$ the basis $\bu$ is generally 
related to $\bu_{\rm c}$ by a unitary transformation 
$\bu=\bu_{\rm c}\bS_{\rm e}$. Then transforming $u$ as 
$u=u_{\rm e}S_{\rm e}$, where the unitary $S_{\rm e}$ is subject to 
${\det}_{\rm w}S_{\rm e}\cdot{\det}_{\rm\bw}\bS_{\rm e}\dg=$ const, 
according to \re{DtD} with \re{UNI} the pair $u_{\rm e}$, $\bu_{\rm c}$ is 
in the same equivalence class as the pair $u$, $\bu$. Analogously for a 
transformed pair $u'$, $\bu'$ we get the equivalent one $u_{\rm e}'$, 
$\bu_{\rm c}$.  Instead of \re{uTs} we then have 
\be
u_{\rm e}'=\T u_{\rm e}\ts,\qquad \bu_{\rm c}=\mbox{const},
\label{uTs1}
\ee
with unitary $\ts$, and instead of \re{D*D} obtain 
\be
{\det}_{\rm w}\ts=e^{i\tilde{\Theta}(\T)}.
\label{D*}
\ee 
Using \re{DtD} with \re{UNI} we thus get the analogon to \re{THE}, 
\be
\tilde{\Theta}(\Id)=\mbox{const},
\label{THe}
\ee
i.e.~again independence of the gauge field at least for $\T=\Id$.

\section{Variational approach}\se

\subsection{General relations}

We define general gauge-field variations of a function $\phi(\U)$ by
\be
\delta\phi(\U)=\frac{\di\phi\big(\U(t)\big)}{\di t}\bigg|_{t=0}\,,\qquad 
\U_{\mu}(t)=e^{t\G_{\mu}^{\rm left}}\U_{\mu}e^{-t\G_{\mu}^{\rm right}},
\label{DEF}
\ee
where $(\U_{\mu})_{n'n}=U_{\mu n}\delta^4_{n',n+\hat{\mu}}$ and 
$(\G_{\mu}^{\rm left/right})_{n'n}=B_{\mu n}^{\rm left/right}\delta^4_{n',n}$.
The special case of gauge transformations then is straightforwardly described 
by
\be
\G_{\mu}^{\rm left}=\G_{\mu}^{\rm right} =\G.
\ee

In the case of gauge transformations we can use the definition 
\re{DEF} and the finite transformation relations to obtain the related
variations explicitly. For operators with ${\cal O}(t)=\T(t)\,
{\cal O}\,\T\dg(t)$ and $\T(t)=e^{t\G}$ this gives 
\be
\delt{\cal O}=[\G,{\cal O}].
\label{DUO}
\ee
In the case $[\T,P_-]\ne0$, $[\T,\bP_+]\ne0$ according to \re{uTs} we have 
for the bases $u(t)=\T(t)u\s(t)$, $\bu(t)=\T(t)\bu\bs(t)$ and obtain
\be
\delt u=\G\,u+u\,\s\dg\,\delt\s,\qquad
\delt\bu=\G\,\bu+\bu\,\bs\dg\,\delt\bs.
\label{DU}
\ee
In the case $[\T,P_-]\ne0$, $\bP_+=$ const according to \re{uTs1} we get 
\be
\delt u_{\rm e}=\G\,u_{\rm e}+u_{\rm e}\,\ts\dg\,\delt\ts,
\qquad\delt\bu_{\rm c}=0.
\label{DU1}
\ee

\subsection{Effective action}

Requiring absence of zero modes of $D$ (and thus also restricting to the 
vacuum sector) the effective action can be considered, for the variation of 
which one gets
\be
\delta\ln{\det}_{\rm\bw w}(\bu\dg Du)=\Tr(P_-D\1\delta D)+
\Tr(\delta u\,u\dg)-\Tr(\delta\bu\,\bu\dg),
\label{EFF}
\ee
in which due to \re{DUO} 
\be
\Tr(P_-D\1\delt D)=\Tr(\G\bP_+)-\Tr(\G P_-).
\label{DE1}
\ee

In the case $[\T,P_-]\ne0$, $[\T,\bP_+]\ne0$ we obtain with \re{DU}
\be
\Tr(\delt u\,u\dg)=\Tr(\G P_-)+\Tr_{\rm w}(\s\dg\,\delt\s),
\ee
\be
\Tr(\delt\bu\,\bu\dg)=\Tr(\G\bP_+)+\Tr_{\rm\bw}(\bs\dg\,\delt\bs),
\ee
and therefore
\be
\delt\ln{\det}_{\rm\bw w}(\bu\dg Du)=\Tr_{\rm w}(\s\dg\,\delt\s)-
\Tr_{\rm\bw}(\bs\dg\,\delt\bs).
\label{DLT}
\ee

In the case $[\T,P_-]\ne0$, $\bP_+=$ const we get with \re{DU1} 
\be
\Tr(\delt u_{\rm e}\,u_{\rm e}\dg)=\Tr_{\rm w}(\ts\dg\,\delt\ts)+\Tr(\G P_-),
\ee
\be
\Tr(\delt\bu_{\rm c}\,\bu_{\rm c}\dg)=0,
\ee
and remembering that $u$, $\bu$ and $u_{\rm e}$, $\bu_{\rm c}$ are in the 
same equivalence class 
\be
\delt\ln{\det}_{\rm\bw w}(\bu\dg Du)=\delt\ln{\det}_{\rm\bw w}
(\bu_{\rm c}\dg Du_{\rm e})=\Tr_{\rm w}(\ts\dg\,\delt\ts)+\Tr(\G\bP_+),
\label{DLT1}
\ee
where now $\Tr(\G\bP_+)$ is constant.

\subsection{Special case of L\"uscher}

L\"uscher \cite{lu98} considers the variation of the effective action 
imposing the Ginsparg-Wilson relation \cite{gi82} $\{\ga,D\}=D\ga D$
and using chiral projections which correspond to the choice $\bG=\Id$ and 
$G=\Id-D$ in \re{GaG}. He assumes $\bP_+$ to be represented by constant
bases so that he is effectively starting from the pair $u_{\rm e}$, 
$\bu_{\rm c}$ of our formulation. 

An important point in L\"uscher's work is the definition of a current 
$j_{\mu n}$ by
\be
\Tr(\delta u_{\rm e}\,u_{\rm e}\dg)=-i\sum_{\mu,n}\mbox{tr}_{\g}
(\eta_{\mu n}j_{\mu n}),\qquad\delta U_{\mu n}=\eta_{\mu n}U_{\mu n},
\label{UE}
\ee
which he requires to transform gauge-covariantly.

His generator is given by $\eta_{\mu n}=B_{\mu,n+\hat{\mu}}^{\rm left}-
U_{\mu n}B_{\mu n}^{\rm right}U_{\mu n}\dg$ in terms of our left and right
generators. We get explicitly
\be
j_{\mu n}=i(U_{\mu n}\rho_{\mu n}+\rho_{\mu n}\dg U_{\mu n}\dg),
\qquad\rho_{\mu n,\alpha'\alpha}=\sum_{j,\sigma}u_{j\sigma}\dg
\frac{\partial u_{\sigma j}\hspace*{7mm}}{\partial U_{\mu n,\alpha\alpha'}}.
\ee
The requirement of gauge-covariance 
$j_{\mu n}'=e^{B_{n+\hat{\mu}}}j_{\mu n}e^{-B_{n+\hat{\mu}}}$ because 
of $U_{\mu n}'=e^{B_{n+\hat{\mu}}}U_{\mu n}e^{-B_n}$ implies that one 
must have 
\be
\rho_{\mu n}'=e^{B_n}\rho_{\mu n}e^{-B_{n+\hat{\mu}}},
\ee
which with \re{uTs1} leads to the condition
\be
\sum_{j,k}\ts_{kj}\dg\frac{\partial\ts_{jk}\hspace*{7mm}}{\partial 
U_{\mu n,\alpha\alpha'}}=0.
\label{II}
\ee
Using \re{II} it follows that
\be
\Tr_{\rm w}(\ts\dg\delta\ts)=0.
\label{DELT}
\ee

Because of $\Tr_{\rm w}(\ts\dg\delta\ts)=\,\delta\ln{\det}_{\rm w}\ts$ it is 
seen that \re{DELT} requires ${\det}_{\rm w}\ts$ to be independent of the 
gauge field.  With \re{D*} we thus obtain
\be
\tilde{\Theta}(\T)=\mbox{const},
\label{THeg}
\ee
i.e.~that \re{THe} extends to all $\T$.  

Since \re{DELT} implies $\Tr_{\rm w}(\ts\dg\delt\ts)=0$ and because 
of the particular form $\bP_+=\h(1+\ga)\Id$ we now get from \re{DLT1} 
\be
\delt\ln{\det}_{\rm\bw w}(\bu\dg Du)=\h\Tr\,\G,
\label{DTD}
\ee
i.e.~a definite result following without any further assumptions. 

Because $\exp(\h\Tr\,\G)$ in \re{DTD} depends only on the gauge 
transformation but not on the gauge field the chiral determinant is 
gauge invariant up to a constant (gauge-field independent) phase factor,
which is $\exp(i\tilde{\Theta}+\h\Tr\,\G)$ as will be confirmed by the 
general analysis in Section 4.3.

\subsection{Both chiral projections non-commuting}

In the case $[\T,P_-]\ne0$, $[\T,\bP_+]\ne0$, starting analogously 
from $\Tr(\delta u\,u\dg)-\Tr(\delta\bu\,\bu\dg)$ as in \re{UE}
from $\Tr(\delta u_{\rm e}\,u_{\rm e}\dg)$, one straightforwardly arrives
at the analogon of \re{DELT},
\be
\Tr_{\rm w}(\s\dg\delta\s)-\Tr_{\rm\bw}(\bs\dg\delta\bs)=0.
\label{DeLT}
\ee
This with \re{D*D} leads to
\be
\Theta(\T)=\mbox{const},
\label{THEg}
\ee
generalizing \re{THE} to all $\T$. With \re{DeLT} we get from \re{DLT}
\be
\delt\ln{\det}_{\rm\bw w}(\bu\dg Du)=0,
\ee
i.e.~again a definite result following without any further assumptions. 
The chiral determinant thus is gauge invariant up to a constant phase 
factor, i.e. up to the factor $\exp(i\Theta)$ as will be confirmed by 
the general analysis in Section 4.2.

\subsection{Discussion}

The definite result \re{DTD} shows that the developments presented in 
Ref.~\cite{lu98} are not relevant for the question of gauge invariance. It 
reveals that whatever the gauge-field dependences of the bases might 
be\footnote{For an observation with respect to general gauge-field 
dependences see the end of Section 4.1.} their gauge-transformation 
properties are such that gauge variations of the effective action in the 
special case there are equal to $\h\Tr\,\G$. (Furthermore, also the aim 
$\delt\ln{\det}_{\rm\bw w} (\bu\dg Du)=0$ in Ref.~\cite{lu98} disagrees 
with the precise result \re{DTD}).

In Ref.~\cite{lu98} a main argument was that without the anomaly cancelation
condition one would be unable to cancel the anomaly term. However, as has
become explicit here this is not true. Indeed, in the case considered there
the basis term being of form $\Tr(\G P_-)$ just compensates the respective 
contribution in the anomaly term \re{DE1} so that one gets cancelation up 
to the irrelevant quantity $\Tr(\G\bP_+)=\h\Tr\,\G$. In the case where both 
chiral projections do not commute with $\T$ the contribution 
$\Tr(\G P_-)-\Tr(\G\bP_+)$ of the bases even fully compensates the anomaly 
term.

Thus in detail the considerations of topological fields (anyway only 
feasible in the Abelian case) and the substantial additional assumptions 
in Ref.~\cite{lu98} turn out to be irrelevant for gauge invariance.

It is to be emphasized in the present context that here as well as in 
Ref.~\cite{lu98} and in the related discussion \cite{go01} one is concerned
with the formulation on the finite lattice and the non-perturbative 
description. Thus one cannot a priori expect to find the same situation as 
in continuum perturbation theory.\footnote{In the limit 
of a perturbation expansion the compensating basis terms vanish so that 
one gets the usual setting of continuum perturbation theory where the
anomaly cancelation condition is needed \cite{ke03}.}

\section{General analysis}\se

\subsection{Equivalence-class requirements}

So far the conditions \re{THeg} and \re{THEg} have emerged as consequences 
of the covariance requirement for L\"uscher's current and have been seen
to prevent the addition of completely arbitrary terms to the gauge variation
of the effective action.
We now turn to the general principle from which these conditions follow. 

In the case where $[\T,P_-]\ne0$ and $[\T,\bP_+]\ne0$ from \re{uTs} with 
\re{D*D},
\be
u'=\T u\s,\qquad\bu'=\T\bu\bs,\qquad
{\det}_{\rm w}\s\cdot{\det}_{\rm\bw}\bs\dg=e^{i\Theta},
\label{UTS}
\ee 
and in the case where $[\T,P_-]\ne0$ and $\bP_+=$ const from 
 and \re{uTs1} with \re{D*}, 
\be
u_{\rm e}'=\T u_{\rm e}\ts,\qquad \bu_{\rm c}=\mbox{const},
\qquad{\det}_{\rm w}\ts=e^{i\tilde{\Theta}},
\label{UTS1}
\ee
it is seen that admitting gauge-field dependence of $\Theta$ and 
of $\tilde{\Theta}$, respectively, means to allow switching to arbitrary 
inequivalent subsets of pairs of bases. Such combinations of gauge 
transformations with transformations to arbitrary different equivalence 
classes of pairs of bases would obviously introduce severe ambiguities. 
Thus to avoid these ambiguities by requiring \re{THEg}, 
\be
\Theta=\mbox{const},
\label{TT}
\ee
and \re{THeg}, 
\be
\tilde{\Theta}=\mbox{const},
\label{TT1}
\ee
respectively, turns out to be appropriate, which also accounts for the fact 
that to describe physics one must restrict to one of the equivalence classes. 

The important observation here is that given an equivalence class of pairs
of bases in this way the equivalence class after the transformation remains 
uniquely determined, which is possible because in the case of gauge 
transformations we have the explicit relations \re{uTs} and \re{uTs1},
respectively. In contrast to this, considering general gauge-field 
dependences, given an equivalence class for one set of fields there is so 
far no criterion determining the equivalence class after a general change 
of the fields.

\subsection{Non-commuting chiral projections}

Considering the transformation of correlation functions in the case where 
$[\T,P_-]\ne0$ and $[\T,\bP_+]\ne0$ using \re{UTS} we obtain
\ba
\langle\psi_{\sigma_1'}'\ldots\psi_{\sigma_R'}'\bar{\psi}_{\bar{\sigma}_1'}'
\ldots\bar{\psi}_{\bar{\sigma}_{\bar{R}}'}'\rangle_{\f}'=\hspace{86mm}
\nonumber\\
e^{i\Theta}\sum_{\sigma_1,\ldots,\sigma_R}\sum_{\bar{\sigma}_1,
\ldots,\bar{\sigma}_{ \bar{R}}}\T_{\sigma_1'\sigma_1}\ldots\T_{\sigma_R'
\sigma_R} \langle\psi_{\sigma_1}\ldots\psi_{\sigma_R}
\bar{\psi}_{\bar{\sigma}_1} \ldots\bar{\psi}_{\bar{\sigma}_{\bar{R}}}
\rangle_{\f}\,\T_{\bar{\sigma}_1\bar{\sigma}_1'}\dg\ldots
\T_{\bar{\sigma}_{\bar{R}}\bar{\sigma}_{\bar{R}}'}\dg.
\label{COV}
\ea
Due to \re{TT} the correlation functions thus turn out to transform 
gauge-covariantly up to a constant phase factor, i.e.~up to the factor 
$e^{i\Theta}$.

\subsection{One constant chiral projection}

In the case where $[\T,P_-]\ne0$ and $\bP_+=$ const we can rewrite 
$\bu_{\rm c}$ as
\be
\bu_{\rm c}=\T\bu_{\rm c}S_{\T}
\label{uS2}
\ee
where $S_{\T}$ because of $[\T,\bP_+]=0$ is unitary. Using this and \re{UTS1} 
we get for the transformation of the correlation functions again the form 
\re{COV} but with $\Theta$ being replaced by $\tilde{\Theta}+\theta_{\T}$ 
where $\theta_{\T}$ is given by
\be
e^{i\theta_{\T}}={\det}_{\rm\bw}S_{\T}\dg
={\det}_{\rm\bw}(\bu_{\rm c}\dg \T\bu_{\rm c}).
\label{ET}
\ee
Since \re{ET} does not depend on in the gauge field and because 
$\tilde{\Theta}$ according to \re{TT1} is also constant the correlation 
functions thus are again seen to show gauge-covariant behavior up to a 
constant phase factor, i.e.~up to the factor $e^{i(\tilde{\Theta}+
\theta_{\T})}$.

To calculate $i\theta_{\T}$ we note that with $[\T,\bP_+]=0$ and 
$\T=e^{\G}$ we get $\bu_{\rm c}\dg \T\bu_{\rm c}=\bu_{\rm c}\dg e^{\G\bP_+}
\bu_{\rm c}$ and the eigenequations $\G\bP_+\bu_j^{\rm d}=
\omega_j\bu_j^{\rm d}$ and $\bP_+\bu_j^{\rm d}=\bu_j^{\rm d}$. With this
we obtain ${\det}_{\rm\bw}(\bu_{\rm c}\dg e^{\G\bP_+}\bu_{\rm c})=
\prod_j e^{\omega_j}=\exp(\Tr(\G\bP_+))$, so that we find $i\theta_{\T}=
\Tr(\G\bP_+)$. For $\bP_+=\h(1+\ga)\Id$ we then have
i$\theta_{\T}= \h\Tr\,\G$.

\section{Conclusions}\se

We have given an unambiguous derivation of the gauge-transformation 
properties in chiral gauge theories on the finite lattices observing that
there are more informations on the bases available which must not  
be ignored. 

We have first considered the subject in terms of variations of the 
effective action In this context we have shown that satisfying the 
covariance requirement for L\"uscher's current the gauge variation leads 
to a definite field-independent quantity without any further assumptions. 
This means that the developments presented in Ref.~\cite{lu98} are 
irrelevant for the question of gauge invariance.

In detail it has become explicit that on the lattice the anomaly term is 
canceled without imposing a respective condition. Thus the considerations 
of topological fields and the substantial additional assumptions in 
Ref.~\cite{lu98} have turned out to be not relevant for gauge invariance.

In our more general analysis we then have pointed out that not allowing 
to combine gauge transformations with arbitrary switching to different 
equivalence classes of pairs of bases is the general principle which also 
implies covariance for L\"uscher's current.

In order to extend the considerations beyond the vacuum sector we 
have investigated the behavior of correlation functions also in the 
presence of zero modes and for any value of the index using finite gauge 
transformations. We have found that fermionic correlation
functions transform gauge-covariantly up to constant phase factors.

\section*{Acknowledgement}

I wish to thank Michael M\"uller-Preussker and his group for their kind
hospitality.

\end{document}